\documentclass[aps,prd,
preprint,
eqsecnum,showpacs,amsmath,nofootinbib]{revtex4-1}

\usepackage{color,graphicx}
\usepackage{amsfonts}
\usepackage{amssymb}

\begin{document}

\title{
Black hole shadow in an asymptotically-flat, stationary, and axisymmetric spacetime: 
The Kerr-Newman and rotating regular black holes
}

\author{Naoki Tsukamoto}\email{tsukamoto@rikkyo.ac.jp}

\affiliation{
School of Physics, Huazhong University of Science and Technology, Wuhan 430074, China 
}

\begin{abstract}
The shadow of a black hole can be one of the strong observational evidences for stationary black holes.
If we see shadows at the center of galaxies, 
we would say whether the observed compact objects are black holes.
In this paper, we consider a formula for the contour of a shadow 
in an asymptotically-flat, stationary, and axisymmetric black hole spacetime. 
We show that the formula is useful for obtaining the contour of the shadow of several black holes 
such as the Kerr-Newman black hole and rotating regular black holes.
Using the formula, we can obtain new examples of the contour of the shadow of rotating black holes  
if assumptions are satisfied.   
\end{abstract}


\maketitle

\section{Introduction}
Recently, LIGO detected three gravitational wave events from binary black hole systems~\cite{Abbott:2016blz,Abbott:2016nmj,Abbott:2017vtc}.
The events showed stellar-mass black holes really exist in our universe.
The physics in strong gravitational field near black holes will be an important topic in not only general relativity but also astronomy.
The black holes are described well by the Kerr black hole solution 
with an Arnowitt-Deser-Misner (ADM) mass $M$ and an angular momentum $J$
which is an exact solution of the Einstein equations.
There, however, remains some possibility for the other black hole solutions 
because of uncertainty in measuring of gravitational waves~\cite{Konoplya:2016pmh}.

It is believed that there are supermassive black holes in the center of galaxies 
and that isolated stationary black holes would be described well by the Kerr solution.
From observations in weak gravitational fields, 
one can estimate the ADM mass $M$ of suppermassive black holes and the distance between the observer 
and the black holes~\cite{Ghez:2008ms,Gillessen:2008qv,Meyer:2012hn,Reid:2014boa,Do:2013upa,Chatzopoulos}.
To get an evidence that the suppermassive compact objects in the centers of galaxies are black holes 
and that they are not the other exotic compact objects predicted by general relativity,  
we should pay attention to phenomena in strong gravitational fields such as 
an optically thin emission region around a black hole~\cite{Falcke:1999pj},
emissions from a geometrically thin accretion disk around a black hole~\cite{Luminet,Fukue,Takahashi:2004xh},
and the shadow made by a black hole~\cite{Bardeen:1973tla}.

In the near future, we may check whether suppermassive black holes in our galaxy and near galaxies can be described by the Kerr black hole.
The Event Horizon Telescope challenges us to measure a shadow made by the suppermassive black hole in the center of our galaxy~\cite{Fish:2016jil,Doeleman:2017nxk}. 
Therefore, the details of the shadows of rotating black holes have been investigated eagerly.  
From an observational viewpoint to check that isolated stationary black holes in nature can be described by the Kerr black hole well, 
the contours of the shadows of dozens rotating black holes have been calculated 
already~\cite{deVries:2000,Takahashi:2005hy,Amarilla:2010zq,Kraniotis:2014paa,Nitta:2011in,Schee:2008fc,Amarilla:2011fx,Yumoto:2012kz,Abdujabbarov:2012bn,Atamurotov:2013dpa,Atamurotov:2013sca,%
Amarilla:2013sj,Li:2013jra,Wei:2013kza,Tsukamoto:2014tja,Grenzebach:2014fha,Papnoi:2014aaa,Wei:2015dua,Abdolrahimi:2015rua,Grenzebach:2015oea,Ghasemi-Nodehi:2015raa,%
Cunha:2015yba,Johannsen:2015hib,Tinchev:2015apf,Shipley:2016omi,Amir:2016cen,Abdujabbarov:2016hnw,Vincent:2016sjq,Dastan:2016vhb,Younsi:2016azx,Tretyakova:2016ale,%
Dastan:2016bfy,Sharif:2016znp,Cunha:2016wzk,Singh:2017vfr,Wang:2017hjl,Amir:2017slq}.
Semianalytic calculations~\cite{Johannsen:2015qca} and
a new numerical method~\cite{Younsi:2016azx} of the contour of the shadow of rotating black holes were investigated.
Hioki and Maeda discussed a relation
between the shape of contour of the shadow and the inclination angle and the spin parameter of the Kerr black hole~\cite{Hioki:2009na}.
Tsupko estimated analytically the spin parameter from the shape of the shadow of the Kerr black hole~\cite{Tsupko:2017rdo}.
A method for distinguishing the Kerr black hole from other rotating black holes with the shape of the shadow 
has been investigated by Tsukamoto \textit{et al.}~\cite{Tsukamoto:2014tja} and by Abdujabbarov \textit{et al.}~\cite{Abdujabbarov:2015xqa}.

The theoretical aspects related to the contour of black hole shadows also have been investigated
since the null geodesics near black holes and other compact objects would determine several important properties of spacetimes.
It was pointed that quasinormal modes~\cite{Vishveshwara:1970zz} 
of a static, spherically symmetric, and asymptotically flat black hole in the eikonal limit 
are determined by the parameters of the unstable circular null geodesics~\cite{Cardoso:2008bp,Mashhoon:1985cya,Hod:2009td}.
A tight relation between quasinormal modes and gravitational lensing near unstable circular null geodesic~\cite{Bozza:2008mi}
was also considered~\cite{Stefanov:2010xz,Wei:2013mda,Raffaelli:2014ola}.
Very recently, a relation between black hole shadows, spacetime instabilities, and fundamental photon orbits which is a generalization of circular photon orbits 
was discussed in Ref.~\cite{Cunha:2017eoe}.

These close links between circular null geodesics, quasinormal modes, gravitational lensing, the shadows, and the other phenomena in strong gravitational fields 
in black hole spacetimes might be valid for only some well-known black holes and one might find counterexamples.
For an example, Konoplya and Stuchlik showed that an expected link between the circular null geodesics and quasinormal modes is broken 
in an asymptotically flat black hole spacetime in the Einstein-Lovelock theory~\cite{Konoplya:2017wot}.
From the theoretical viewpoint, it would be worth to investigate the details of the null geodesic near less-known black holes and of their shadows.

In this paper, 
we investigate an simple analytic formula for the contour of the shadow of rotating black holes
and we apply the formula to two new examples of rotating black holes
and then we examine several known results with the formula.
We emphasize that the formula would help to categorize the shadow of dozens rotating black holes,
and that it can describe new examples of the contours of the shadows of rotating black holes.
 
This paper is organized as follows. 
In Sec.~II, we introduce a line element describing a rotating black hole spacetime. 
In Sec.~III, we obtain null geodesic equations in the rotating black hole spacetime. 
In Sec.~IV, we investigate a formula for the contour of the shadow of the rotating black hole 
and we apply the formula to two examples of rotating black holes.  
In Sec.~V, we examine several known results by using the formula.
In Sec.~VI, we summarize our result.
In this paper, we use the units in which the light speed and Newton's constant are unity.

\section{A line element in a rotating black hole spacetime}
A Newman-Janis algorithm~\cite{Newman:1965tw} generates a stationary and axisymmetric black hole spacetime  
from a static and spherical black hole spacetime. 
The Newman-Janis algorithm was investigated to find the Kerr-Newman solution 
is an exact solution of the Einstein-Maxwell equations~\cite{Newman:1965my,Newman:1965tw} 
from the Reissner-Nordstr\"om solution.
Recently, the Newman-Janis algorithm has been eagerly applied for regular black hole spacetime to obtain rotating regular black hole metrics.

We consider a line element in an asymptotically-flat, stationary, and axisymmetric black hole spacetime, in the Boyer-Lindquist coordinates,
\begin{eqnarray}\label{eq:line_1}
ds^{2}
&=& -\frac{\rho^{2}\Delta}{\Sigma}dt^{2} +\frac{\Sigma \sin^{2}\theta}{\rho^{2}} \left[ d\phi-\frac{a(r^{2}+a^{2}-\Delta)}{\Sigma}dt \right]^{2} \nonumber\\
&&+\frac{\rho^{2}}{\Delta}dr^{2} +\rho^{2}d\theta^{2} \nonumber\\
&=&-\left(1-\frac{2m(r)r}{\rho^{2}} \right)dt^{2}-\frac{4m(r)ar\sin^{2}\theta}{\rho^{2}}d\phi dt \nonumber\\
&&+\left( r^{2}+a^{2}+ \frac{2m(r)a^{2}r\sin^{2}\theta}{\rho^{2}} \right) \sin^{2}\theta d\phi^{2} \nonumber\\
&&+\frac{\rho^{2}}{\Delta}dr^{2} +\rho^{2}d\theta^{2},
\end{eqnarray}
where
\begin{eqnarray}
\rho^{2}&\equiv&r^{2}+a^{2}\cos^{2} \theta , \nonumber\\
\Delta(r)&\equiv&r^{2}-2m(r)r+a^{2}, \nonumber\\
\Sigma&\equiv&(r^{2}+a^{2})^{2}-a^{2}\Delta(r) \sin^{2}\theta
\end{eqnarray}
and where $a$ is a spin parameter defined as $a\equiv J/M$ 
and $M$ and $J$ are the ADM mass and the angular momentum of the black hole, respectively, 
and $m(r)$ is a function with respect to the radial coordinate satisfying $m(r) \rightarrow M$ as $r \rightarrow \infty$.
The line element~(\ref{eq:line_1}) can be obtained 
when we apply the Newman-Janis algorithm for an asymptotically-flat, static, and spherical black hole spacetime with a line element~\cite{Bambi:2013ufa},
\begin{eqnarray}\label{eq:line_0}
ds^{2}
&=&-\left(1-\frac{2m(r)}{r} \right)dt^{2} +\left(1-\frac{2m(r)}{r} \right)^{-1}dr^{2}\nonumber\\
&&+r^{2} ( d\theta^{2} +\sin^{2}\theta d\phi^{2}).
\end{eqnarray}

We assume the existence of an event horizon at $r=r_{+}$.
In other words, 
we assume that an equation $\Delta(r)=0$ has one or more positive solutions 
and that $r=r_{+}$ is the largest positive solution among them.
We also assume that $m(r)$ is regular in a range $r\geq r_{+}$.
In Secs.~IV and V, we show that several examples of the function $m(r)$. 

The components of the inverse metric are given by
\begin{eqnarray}
&&g^{tt}=-\frac{\Sigma}{\rho^{2}\Delta},\\
&&g^{t\phi}=-\frac{2m(r)ar}{\rho^{2}\Delta},\\
&&g^{\phi\phi}=\frac{\Delta -a^{2}\sin^{2}\theta}{\rho^{2}\Delta \sin^{2}\theta},\\
&&g^{rr}=\frac{\Delta}{\rho^{2}},\\
&&g^{\theta\theta}=\frac{1}{\rho^{2}}.
\end{eqnarray}

\section{Null Geodesic Equation}
In this section, we investigate the Hamilton-Jacobi method for the motion of a photon in the asymptotically-flat, stationary and axisymmetric black hole spacetime. 
We define the action $S=S(\lambda, x^{\mu})$ as a function of the coordinates $x^{\mu}$ and the parameter $\lambda$.
The Hamilton-Jacobi equation is obtained by
\begin{eqnarray}
\frac{\partial S}{\partial \lambda} +H =0,
\end{eqnarray}
where $H\equiv g_{\mu \nu}p^{\mu}p^{\nu}/2$ is the Hamiltonian of the photon motion
and $p_{\mu}$ is the conjugate momentum of the photon given by 
\begin{eqnarray}
p_{\mu}\equiv \frac{\partial S}{\partial x^{\mu}}.
\end{eqnarray}
We can write the action $S$ in the following form with the cyclic coordinates $t$ and $\phi$:
\begin{eqnarray}
S=\frac{1}{2}\mu^{2}\lambda -Et+L \phi +S_{r}(r) +S_{\theta}(\theta), 
\end{eqnarray}
where  the conserved energy $E\equiv -p_{t}$, the conserved angular momentum $L\equiv p_{\phi}$ and the mass $\mu\equiv -p_{\mu}p^{\mu}=0$ of the photon 
are constant along the geodesic
and $S_{r}(r)$ and $S_{\theta}(\theta)$ are functions of the coordinates $r$ and $\theta$, respectively.
We can rewrite the Hamilton-Jacobi equation in
\begin{eqnarray}
&&-\Delta \left( \frac{dS_{r}}{dr} \right)^{2} +\frac{\left[ (r^{2}+a^{2})E-aL \right]^{2}}{\Delta} \nonumber\\
&&=\left( \frac{dS_{\theta}}{d\theta} \right)^{2} +\frac{(L-aE\sin^{2}\theta)^{2}}{\sin^{2}\theta}.
\end{eqnarray}
Both sides of this equation are constant. 
We can divide the Hamilton-Jacobi equation into two equations with respect to $r$ and $\theta$:
\begin{eqnarray}
\mathcal{K}=-\Delta \left( \frac{dS_{r}}{dr} \right)^{2} +\frac{\left[ (r^{2}+a^{2})E-aL \right]^{2}}{\Delta}
\end{eqnarray}
and
\begin{eqnarray}
\mathcal{K}=\left( \frac{dS_{\theta}}{d\theta} \right)^{2} +\frac{(L-aE\sin^{2}\theta)^{2}}{\sin^{2}\theta},
\end{eqnarray}
where $\mathcal{K}$ is a constant.

From 
\begin{eqnarray}
\frac{dx^{\mu}}{d\lambda}=p^{\mu}=g^{\mu \nu} p_{\nu},
\end{eqnarray}
we obtain 
\begin{eqnarray}
&&\rho^{2}\frac{dt}{d\lambda}=-a(aE\sin^{2}\theta-L)+\frac{(r^{2}+a^{2})P(r)}{\Delta(r)}, \\
&&\rho^{2}\frac{dr}{d\lambda}=\sigma_{r} \sqrt{R(r)}, \\ \label{eq:Theta_motion}
&&\rho^{2}\frac{d\theta}{d\lambda}=\sigma_{\theta} \sqrt{\Theta(\theta)}, \\
&&\rho^{2}\frac{d\phi}{d\lambda}=-\left( aE-\frac{L}{\sin^{2}\theta} \right)+\frac{aP(r)}{\Delta(r)},
\end{eqnarray}
where
\begin{eqnarray}
&&P(r)\equiv E(r^{2}+a^{2})-aL,\\\label{eq:R_define}
&&R(r)\equiv P(r)^{2}-\Delta(r) \left[ (L-aE)^{2}+\mathcal{Q} \right],\\
&&\Theta(\theta) \equiv \mathcal{Q}+\cos^{2}\theta \left( a^{2}E^{2}-\frac{L^{2}}{\sin^{2}\theta} \right)
\end{eqnarray}
and where $\sigma_{r}=\pm 1$ and $\sigma_{\theta}=\pm 1$ are independent 
and $\mathcal{Q}$ is the Carter constant defined by $\mathcal{Q}\equiv \mathcal{K}-(L-aE)^{2}$.
$R(r)$ and $\Theta(\theta)$ should be non-negative for the photon motion.

From $\Theta(\theta) \geq 0$, we obtain
\begin{equation}\label{eq:Theta_inequality}
\frac{\Theta(\theta)}{E^{2}} 
=\eta +(a-\xi)^{2} -\left( a\sin \theta -\frac{\xi}{\sin \theta} \right)^{2} \geq 0,
\end{equation}
where $\eta \equiv \mathcal{Q}/E^{2}$ and $\xi \equiv L/E$.

Equation (\ref{eq:R_define}) can be rewritten in
\begin{equation}
\frac{R(r)}{E^{2}}=r^{4}+ ( a^{2}-\xi^{2}-\eta )r^{2} +2m(r) \left[ (\xi-a)^{2}+\eta \right]r -a^{2}\eta
\end{equation}
and the derivative with respective to the radial coordinate $r$ is given by
\begin{equation}
\frac{R'}{E^{2}}=4r^{3}+2(a^{2}-\xi^{2}-\eta)r +2m(r) \left[ (\xi-a)^{2} +\eta \right] f(r),
\end{equation}
where $'$ denotes the differentiation with respect to the radial coordinate $r$
and where $f(r)$ is defined as
\begin{equation}
f(r)\equiv 1+\frac{m'r}{m}.
\end{equation}

\section{The shadow of the rotating black hole}
In this section, we consider a formula for the contour of the shadow of the rotating black hole and apply it for new examples.

\subsection{A formula for the black hole shadow}
We give a formula for the contour of the shadow of the rotating black hole.
We assume that the rotating black hole spacetime has unstable circular null orbits satisfying 
\begin{eqnarray}\label{eq:circular1}
\left.R(r) \right|_{r=r_{0}}=\left. R'(r)\right|_{r=r_{0}}=0, 
\end{eqnarray}
and 
\begin{eqnarray}
\left.R''(r) \right|_{r=r_{0}} > 0,
\end{eqnarray}
where $r_0$ is the radius of the unstable circular null orbits. 
We also assume 
\begin{eqnarray}
r_{+} \leq r_{0}.
\end{eqnarray}

From Eq. (\ref{eq:circular1}), we obtain 
\begin{eqnarray}\label{eq:R_vanish1}
&&(4-f_{0})r_{0}^{4} +(2-f_{0})r_{0}^{2}a^{2} -[(2-f_{0})r_{0}^{2}-f_{0}a^{2}]\eta \nonumber\\
&&=(2-f_{0})r_{0}^{2}\xi^{2}
\end{eqnarray}
and
\begin{eqnarray}\label{eq:R_vanish2}
&&r_{0}^{4}-(2-f_{0})m_{0}a^{2}r_{0}+[a^{2}-(2-f_{0})m_{0}r_{0}]\eta \nonumber\\
&&=(2-f_{0})m_{0}r_{0} ( \xi^{2}-2a\xi ),
\end{eqnarray}
where $m_{0}$ and $f_{0}$ are defined as $m_{0}\equiv m(r_{0})$ and $f_{0}\equiv f(r_{0})$, respectively. 
From Eqs. (\ref{eq:R_vanish1}) and (\ref{eq:R_vanish2}), we obtain a quadratic equation with respective to $\xi$ as 
\begin{eqnarray}\label{eq:quadratic}
&&a^{2}(r_{0}-f_{0}m_{0})\xi^{2}-2am_{0}[(2-f_{0})r_{0}^{2}-f_{0}a^{2}]\xi  \nonumber\\
&&-r_{0}^{5} +(4-f_{0})m_{0}r_{0}^{4} -2a^{2}r_{0}^{3} \nonumber\\
&&+2a^{2}m_{0}(2-f_{0})r_{0}^{2} -a^{4}r_{0} -a^{4}m_{0}f_{0}=0.
\end{eqnarray}
Equation~(\ref{eq:quadratic}) has two real solutions $\xi=\xi_{\pm}$ given by
\begin{eqnarray}
\xi_{\pm}\equiv \frac{m_{0}\left[ \left( 2-f_{0} \right) r_{0}^{2}- f_{0} a^{2} \right] \pm r_{0}\Delta_{0}}{a(r_{0}-f_{0}m_{0})},
\end{eqnarray} 
where $\Delta_0$ is defined as $\Delta_0 \equiv \Delta(r_0)$.

We can simplify the solution $\xi_{+}$ as 
\begin{eqnarray}
\xi_{+}=\frac{r_{0}^{2}+a^{2}}{a},
\end{eqnarray}
and, by using Eq.~(\ref{eq:R_vanish1}), we obtain 
\begin{eqnarray}
\eta 
=\eta_{+}
\equiv -\frac{r_{0}^{4}}{a^{2}}
=-(a-\xi_{+})^{2}.
\end{eqnarray}
We notice that $\xi_{+}$ and $\eta_{+}$ are the same as the Kerr black hole case~\cite{Bardeen:1973tla}.
From Eqs.~(\ref{eq:Theta_motion}) and (\ref{eq:Theta_inequality}),
only $\theta=\theta_{+}$, where $\theta_{+}$ is a constant satisfying $\xi_{+}=a\sin^{2}\theta_{+}$,  
is permitted in this case.
Thus, the solution $\xi=\xi_{+}$ must be rejected for our purpose to describe the black hole shadow. 

We choose the solution $\xi=\xi_{-}$, where 
\begin{eqnarray}\label{eq:xi-}
\xi_{-}\equiv \frac{4m_{0}r_{0}^{2}-(r+f_{0}m_{0})(r_{0}^{2}+a^{2})}{a(r_{0}-f_{0}m_{0})}
\end{eqnarray} 
since we are interested in the shadow of the black hole.
From Eq.~(\ref{eq:R_vanish1}), we get $\eta=\eta_{-}$, where 
\begin{equation}\label{eq:eta-}
\eta_{-}
\equiv \frac{r_{0}^{3}\left\{  4(2-f_{0})a^{2}m_{0}-r_{0}\left[r_{0}-(4-f_{0})m_{0}\right]^{2} \right\}}{a^{2}(r_{0}-f_{0}m_{0})^{2}}.
\end{equation}

We consider an asymptotically-flat, stationary, and axisymmetric black hole 
seen by an observer at a large distance from the black hole along an inclination angle $\theta_i$.
The contour of the shadow of the black hole can be expressed 
by celestial coordinates $\alpha$ and $\beta$ in a small part of the celestial sphere of the observer~\cite{Bardeen:1973tla}.
The celestial coordinates $\alpha$ and $\beta$ are obtained by 
\begin{eqnarray}\label{eq:alpha}
\alpha = \frac{\xi_-}{\sin \theta_i}
\end{eqnarray}
and
\begin{equation}\label{eq:beta}
\beta =\sigma_{\theta} \sqrt{ \eta_- +(a-\xi_-)^{2} -\left( a\sin \theta_i -\frac{\xi_-}{\sin \theta_i} \right)^{2} }.
\end{equation}
See Appendix~A for the calculation of the celestial coordinates $\alpha$ and $\beta$.
If we obtain a specific form of the function $m(r)$, from Eqs.~(\ref{eq:xi-}) and (\ref{eq:eta-}), 
we can obtain $\xi_-$ and $\eta_-$.
Then, by using $\alpha$ and $\beta$,
we can draw the contour of the shadow of the black hole with the inclination angle $\theta_i$.

\subsection{Application}
We apply the formula for two new examples of black holes to check that it works well.

\subsubsection{Rotating black hole with $m(r)=2M/(1+e^{s/r})$}
First we consider a rotating black hole applied the Newman-Janis algorithm 
with 
\begin{eqnarray}\label{eq:ms}
m(r)=\frac{2M}{1+e^{s/r}} 
\end{eqnarray}
suggested in Refs.~\cite{AyonBeato:1999rg,Balart:2014cga}.
Here $s$ is a positive constant.
$f_0$ is given by
\begin{eqnarray}\label{eq:f_0s}
f_0=\frac{r_0+(r_0+s)e^{s/r_0}}{r_0(1+e^{s/r_0})}.
\end{eqnarray}
There is a boundary of parameters $a$ and $s$ for existence of the event horizon shown in Fig.~1.
The maximum value of $a/M$ is $1$ for $s/M = 0$, it decreases as $s/M$ increases, and it is $0$ for $s/M \sim 1.114$.
\begin{figure}[htbp]
\begin{center}
\includegraphics[width=85mm]{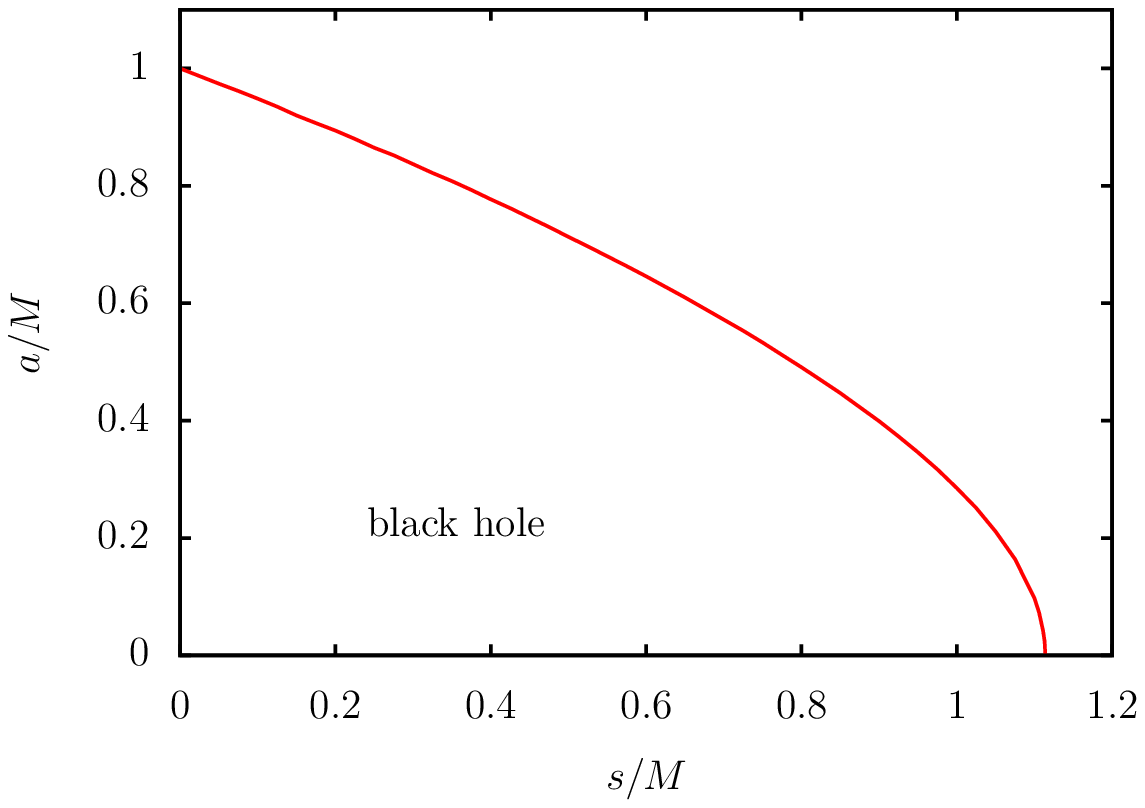}
\end{center}
\caption{The boundary of parameters $a$ and $s$ for the black hole with $m(r)=2M/(1+e^{s/r})$.
For existence of the event horizon, $a$ and $s$ should be smaller than ones on the boundary.}
\end{figure}
From Eqs.~(\ref{eq:xi-}) - (\ref{eq:f_0s}), 
the shadow of the black hole is obtained as shown in Fig.~2. 
\begin{figure}[htbp]
\begin{center}
\includegraphics[width=85mm]{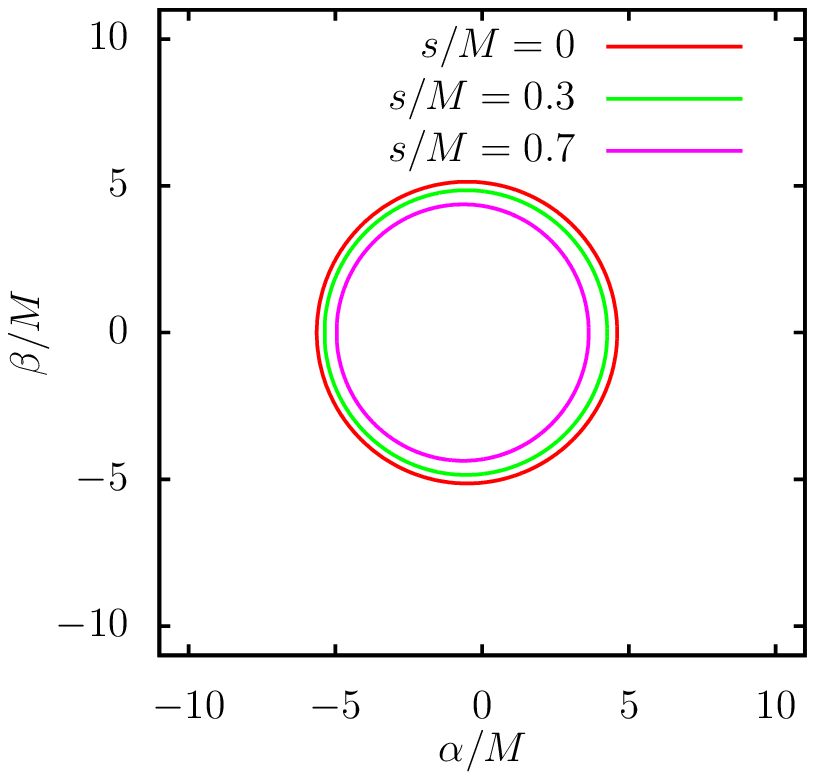}
\includegraphics[width=85mm]{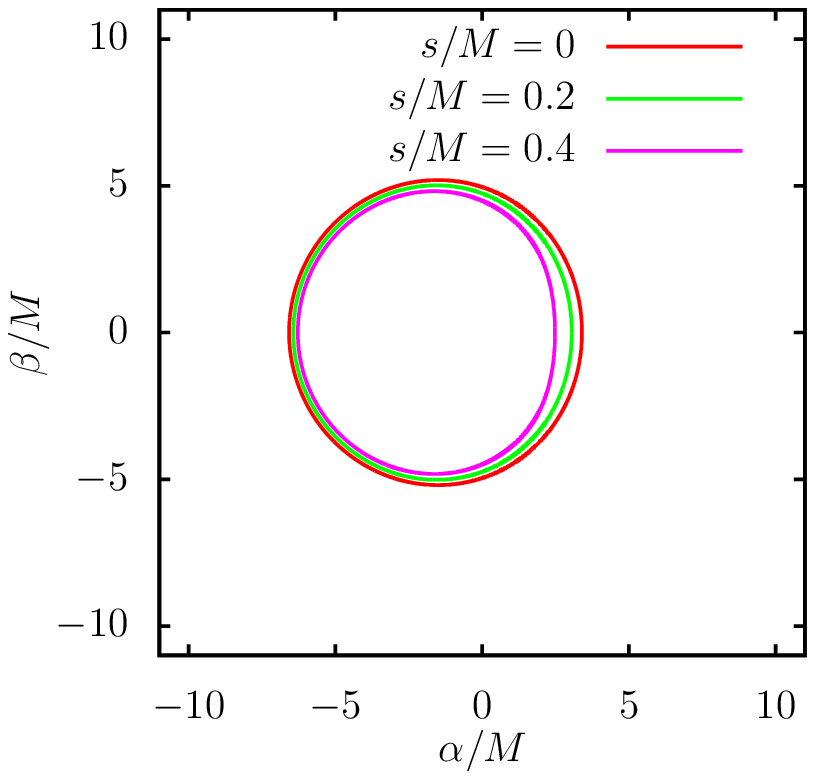}
\end{center}
\caption{The contour of shadow of the black hole with $m(r)=2M/(1+e^{s/r})$.
Top: Outer (red), middle (green), and inner (magenta) circles are the contours of the shadows of the black hole 
with $s/M=0$, $0.3$, and $0.7$, respectively. The spin parameter is $a/M=0.5$ and the inclination angle is $\theta_i=\pi/6$.
Bottom: Outer (red), middle (green), and inner (magenta) circles are the contours of the shadows of the black hole 
with $s/M=0$, $0.2$, and $0.4$, respectively. The spin parameter is $a/M=0.75$ and the inclination angle is $\theta_i=\pi/2$.
}
\end{figure}

\subsubsection{Rotating black hole with $m(r)=4Me^{u/\sqrt{r}}/\left( 1+e^{u/\sqrt{r}}\right)^2$}
Next we consider a rotating black hole applied the Newman-Janis algorithm with 
\begin{eqnarray}\label{eq:mu}
m(r)=\frac{4Me^{u/\sqrt{r}}}{\left( 1+e^{u/\sqrt{r}}\right)^2}
\end{eqnarray}
where $u$ is a positive constant which is suggested in Ref.~\cite{Balart:2014cga}.
$f_0$ is obtained as 
\begin{eqnarray}\label{eq:f_0u}
f_0=1-\frac{(1-e^{u/\sqrt{r_0}})u}{2(1+e^{u/\sqrt{r_0}})\sqrt{r_0}}.
\end{eqnarray}
Figure~3 shows a boundary of parameters $a$ and $u$ for existence of the event horizon.
\begin{figure}[htbp]
\begin{center}
\includegraphics[width=85mm]{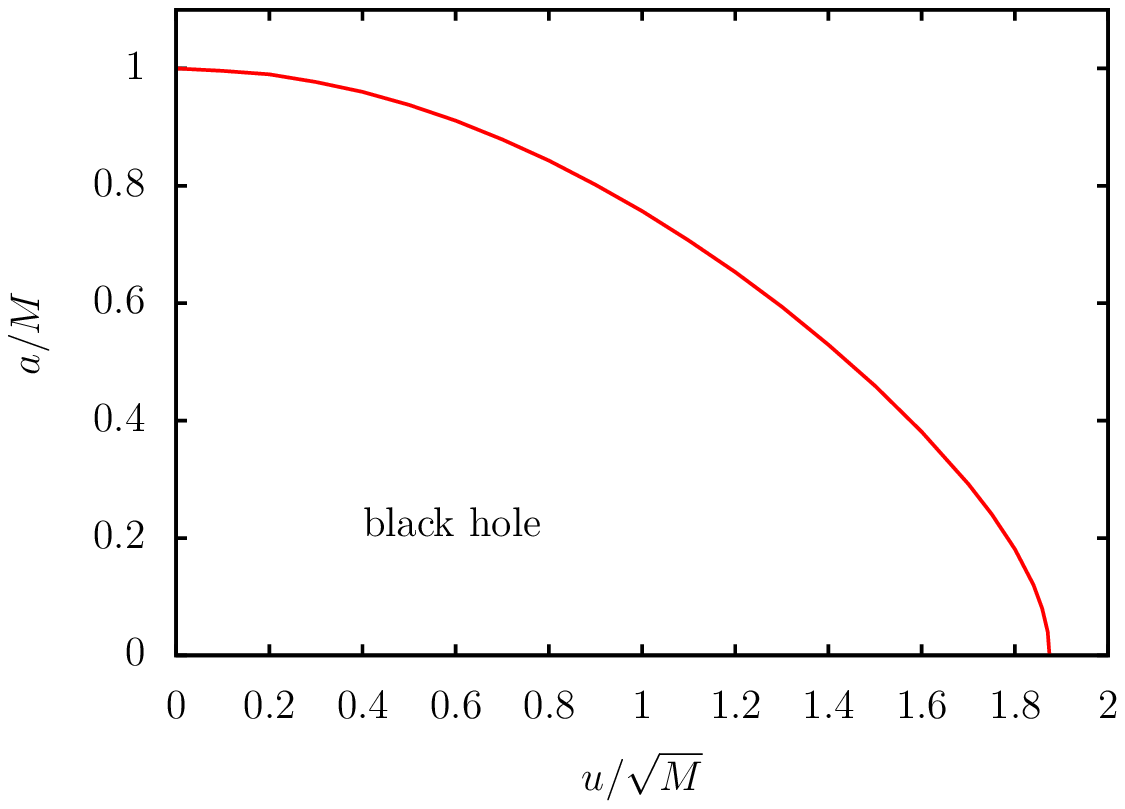}
\end{center}
\caption{The boundary of parameters $a$ and $u$ for the black hole with $m(r)=4Me^{u/\sqrt{r}}/\left( 1+e^{u/\sqrt{r}}\right)^2$.
$a$ and $u$ for the black hole should be smaller than ones on the boundary.}
\end{figure}
The maximum value of $a/M$ is unity for $u/\sqrt{M}= 0$, decreases as $u/\sqrt{M}$ increases, and vanishes for $u/\sqrt{M} \sim 1.874$.
From Eqs.~(\ref{eq:xi-}) - (\ref{eq:beta}), (\ref{eq:mu}), and (\ref{eq:f_0u}), 
the contour of the shadow of the black hole is obtained and it is shown in Fig.~4. 
\begin{figure}[htbp]
\begin{center}
\includegraphics[width=85mm]{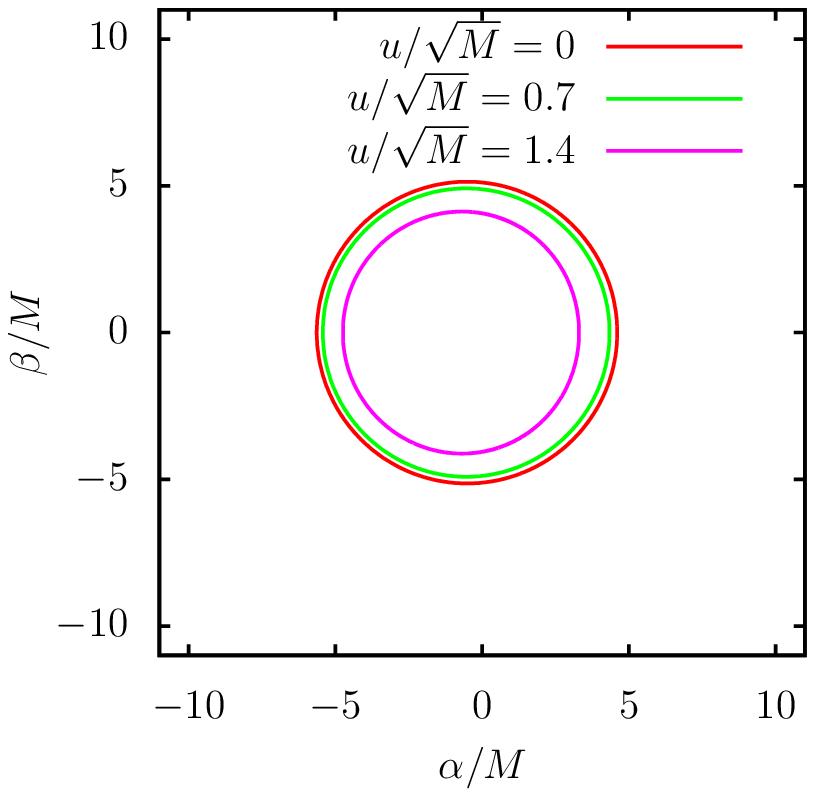}
\includegraphics[width=85mm]{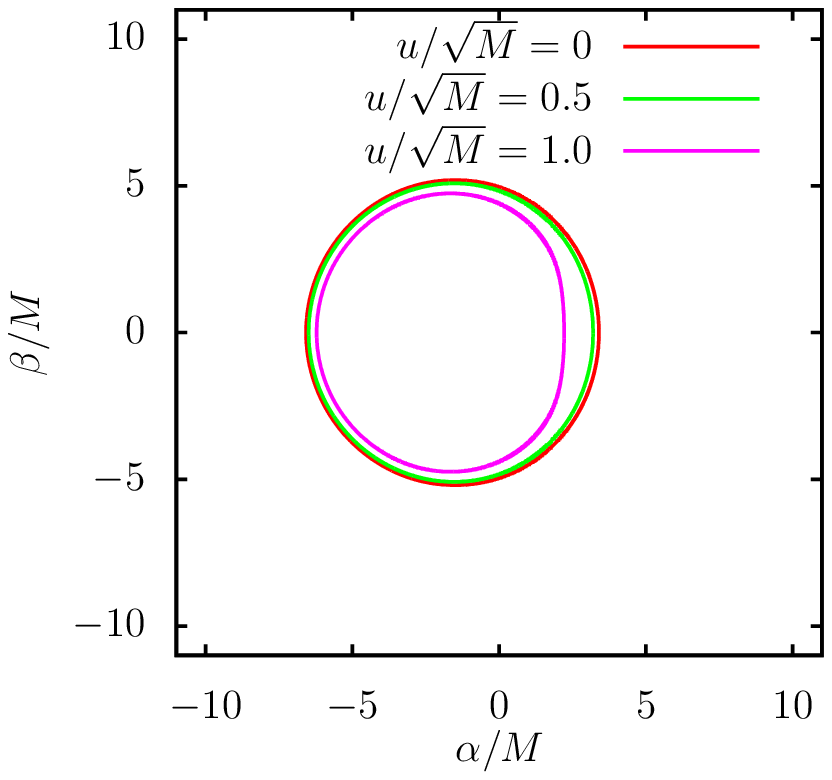}
\end{center}
\caption{The contour of shadow of the black hole with $m(r)=4Me^{u/\sqrt{r}}/\left( 1+e^{u/\sqrt{r}}\right)^2$.
Top: Outer (red), middle (green), and inner (magenta) circles are the contours of the shadows of the black hole 
with $u/\sqrt{M}=0$, $0.7$, and $1.4$, respectively. The spin parameter is $a/M=0.5$ and the inclination angle is $\theta_i=\pi/6$.
Bottom: Outer (red), middle (green), and inner (magenta) circles are the contours of the shadows of the black hole 
with $u/\sqrt{M}=0$, $0.5$, and $1.0$, respectively. The spin parameter is $a/M=0.75$ and the inclination angle is $\theta_i=\pi/2$.
}
\end{figure}

\section{Examination of known results}
In this section, we examine that Eqs.~(\ref{eq:xi-}) and~(\ref{eq:eta-}) recovers 
several known results of rotating black holes such as the Kerr-Newman black hole and rotating regular black holes.

\subsection{The Kerr-Newman black hole}
The Kerr-Newman solution is an exact solution of the Einstein-Maxwell equations~\cite{Newman:1965my,Newman:1965tw}.
The function $m(r)$ of the Kerr-Newman spacetime with the electrical charge $Q$ is given by 
\begin{equation}
m(r)=M-\frac{Q^{2}}{2r}.
\end{equation}
The Kerr-Newman solution is a black hole solution when $M^2-a^2-Q^2 \geq 0$ is satisfied.
The contour of the shadow of the Kerr-Newman black hole was investigated in Ref.~\cite{deVries:2000,Takahashi:2005hy,Kraniotis:2014paa}.
From Eqs.~(\ref{eq:xi-}) and~(\ref{eq:eta-}), $\xi_-$ and $\eta_-$ are given by
\begin{equation}\label{eq:xi-1}
\xi_-=\frac{2r_0(2Mr_0-Q^2)-(r_0+M)(r_0^2+a^2)}{a(r_0-M)}
\end{equation}
and
\begin{equation}\label{eq:eta-1}
\eta_-
=\frac{r_0^2\left\{ 4a^2(Mr_0-Q^2)-\left[r_0(r_0-3M)+2Q^2 \right]^2 \right\}}{a^{2}(r_0-M)^{2}},
\end{equation}
respectively.
Equations~(\ref{eq:xi-1}) and~(\ref{eq:eta-1}) are equal to 
Eqs. (51) and (52) in Ref.~\cite{deVries:2000}
and  Eqs. (6) and (7) in Ref.~\cite{Takahashi:2005hy}.

When the electrical charge $Q$ vanishes, 
the spacetime is the Kerr black hole spacetime~\cite{Kerr:1963ud} and 
$\xi_-$ and $\eta_-$ become
\begin{equation}\label{eq:xi-1b}
\xi_-=\frac{4Mr_0^2-(r_0+M)(r_0^2+a^2)}{a(r_0-M)}
\end{equation}
and
\begin{equation}\label{eq:eta-1b}
\eta_-
=\frac{r_0^3\left[ 4a^2M- r_0(r_0-3M)^2 \right]}{a^{2}(r_0-M)^{2}},
\end{equation}
respectively.
Equations~(\ref{eq:xi-1b}) and~(\ref{eq:eta-1b}) are equal to Eqs. (48) and (49) obtained by Bardeen~\cite{Bardeen:1973tla}.

\subsection{A braneworld black hole with a tidal charge}
A rotating black hole localized on a 3-brane in the Randall-Sundrum braneworld~\cite{Randall:1999ee}
was considered by Aliev and Gumrukcuoglu~\cite{Aliev:2005bi}.
The function $m(r)$ of the braneworld black hole with a tidal charge $b$ is given by 
\begin{equation}
m(r)=M-\frac{b}{2r}.
\end{equation}
When $M^2-a^2-b \geq 0$ is satisfied, the spacetime is a black hole spacetime.
The line element is the same to the one of the Kerr-Newman black hole spacetime if the tidal charge is nonnegative
while it is not if the tidal charge is negative. 
Notice that the spin parameter $a$ can be larger than the ADM mass $M$ 
when the tidal charge $b$ is negative~\cite{Aliev:2005bi}. 

The contour of the shadow of the rotating braneworld black hole with the tidal charge $b$ was investigated in Refs.~\cite{Schee:2008fc,Amarilla:2011fx}.
From Eqs.~(\ref{eq:xi-}) and~(\ref{eq:eta-}), we obtain $\xi_-$ and $\eta_-$ as
\begin{equation}\label{eq:xi-2}
\xi_-=\frac{2r_0(2Mr_0-b)-(r_0+M)(r_0^2+a^2)}{a(r_0-M)}
\end{equation}
and
\begin{equation}\label{eq:eta-2}
\eta_-
=\frac{r_0^2\left\{ 4a^2(Mr_0-b)-\left[r_0(r_0-3M)+2b \right]^2 \right\}}{a^{2}(r_0-M)^{2}},
\end{equation}
respectively.
We notice that Eqs.~(\ref{eq:xi-2}) and~(\ref{eq:eta-2}) are the same as Eqs.~(11) and (12) in Ref.~\cite{Amarilla:2011fx}.

\subsection{Rotating regular black hole}
Recently, rotating regular black hole metrics were suggested eagerly.

\subsubsection{Rotating Bardeen black hole}
Applying the Newman-Janis algorithm to the Bardeen black hole~\cite{Bardeen:1968,Borde:1996df}, 
a rotating regular black hole metric was obtained in Ref.~\cite{Bambi:2013ufa}.
The function $m(r)$ is given by
\begin{equation}
m(r)=M\left( \frac{r^{2}}{r^{2}+c^{2}} \right)^{\frac{3}{2}} 
\end{equation}
where $c$ is the monopole charge of a self-gravitating magnetic field~\cite{AyonBeato:2000zs}.

The shadow of the rotating Bardeen black hole was investigated in Refs.~\cite{Li:2013jra,Tsukamoto:2014tja,Abdujabbarov:2016hnw}.
From Eqs.~(\ref{eq:xi-}) and~(\ref{eq:eta-}), we obtain $\xi_-$ and $\eta_-$ as
\begin{eqnarray}\label{eq:xi-8}
\xi_-=\frac{1}{a\left[(r_0^2+c^2)^\frac{5}{2}-Mr_0^2(r_0^2+4c^2)\right]} \left\{ 4Mr_0^4(r_0^2+c^2) \right. \nonumber\\
\left. -\left[(r_0^2+c^2)^\frac{5}{2}+Mr^2_0(r_0^2+4c^2) \right](r_0^2+a^2) \right\} \nonumber\\
\end{eqnarray}
and 
\begin{eqnarray}\label{eq:eta-8}
\eta_-
&=&
\frac{r_0^{4}}{a^2\left[(r_0^2+c^2)^\frac{5}{2}-Mr_0^2(r_0^2+4c^2)\right]^2} \nonumber\\
&&\times \left\{ 4Ma^{2}(r_0^2+c^2)^\frac{5}{2}(r_0^2-2c^2)\right. \nonumber\\
&& \left.-\left[ (r_0^2+c^2)^\frac{5}{2}-3Mr_0^4 \right]^{2} \right\}, 
\end{eqnarray}
respectively.
Equations~(\ref{eq:xi-8}) and (\ref{eq:eta-8})~ are the same as Eq.~(2.19) in Ref.~\cite{Tsukamoto:2014tja} 
and Eqs.~(41) and (42) in Ref~\cite{Abdujabbarov:2016hnw}.
We comment on Ref.~\cite{Li:2013jra} in Appendix~B.

\subsubsection{Rotating Hayward black hole}
Applying the Newman-Janis algorithm to a regular black hole considered by Hayward~\cite{Hayward:2005gi}, 
a rotating regular black hole metric was obtained in Ref.~\cite{Bambi:2013ufa}.
The function $m(r)$ is given by
\begin{equation}
m(r)=\frac{Mr^{3}}{r^{3}+g^{3}} 
\end{equation}
where $g$ is a constant.

The contour of shadow of the rotating Hayward black hole was considered in Refs.~\cite{Li:2013jra,Abdujabbarov:2016hnw}.
From Eqs.~(\ref{eq:xi-}) and~(\ref{eq:eta-}), $\xi_-$ and $\eta_-$ are given by
\begin{eqnarray}\label{eq:xi-3}
\xi_-=\frac{1}{a\left[(r_0^3+g^3)^2-Mr_0^2(r_0^3+4g^3)\right]} \left\{ 4Mr_0^4(r_0^3+g^3) \right. \nonumber\\
\left. -\left[(r_0^3+g^3)^2+Mr_0^2(r_0^3+4g^3) \right](r_0^2+a^2) \right\} \nonumber\\
\end{eqnarray}
and 
\begin{eqnarray}\label{eq:eta-3}
\eta_-
&=&
\frac{r_0^4}{a^2\left[(r_0^3+g^3)^2-Mr_0^2(r_0^3+4g^3)\right]^2} \nonumber\\
&&\times \left\{ 4Ma^{2}(r_0^3+g^3)^2(r_0^3-2g^3) \right. \nonumber\\
&&\left. -\left[ (r_0^3+g^3)^2-3Mr_0^5 \right]^{2} \right\}, 
\end{eqnarray}
respectively.
Equations~(\ref{eq:xi-3}) and (\ref{eq:eta-3}) are the same as Eqs.~(41) and (42) in Ref~\cite{Abdujabbarov:2016hnw}.
See also Appendix~B.

\subsubsection{A rotating black hole considered by Ghosh~\cite{Ghosh:2014pba}}
Ghosh applied the Newman-Janis algorithm to one of regular black holes suggested by Balart and Vagenas~\cite{Balart:2014cga}
and obtained a rotating regular black hole metric~\cite{Ghosh:2014pba}.
The function $m(r)$ in Ref.~\cite{Ghosh:2014pba} is 
\begin{eqnarray}
m(r)=Me^{-\frac{h}{r}}, 
\end{eqnarray}
where $h$ is constant.

The contour of the shadow of the black hole was calculated by Amir and Ghosh~\cite{Amir:2016cen}.
From Eqs.~(\ref{eq:xi-}) and~(\ref{eq:eta-}), we obtain $\xi_-$ and $\eta_-$ as
\begin{equation}\label{eq:xi-4}
\xi_-=\frac{4Mr_0^3e^{-\frac{h}{r_0}}-\left[r_0^2+M(r_0+h)e^{-\frac{h}{r_0}}\right](r_0^{2}+a^{2})}{a\left[r_0^{2}-M(r_0+h)e^{-\frac{h}{r_0}}\right]}
\end{equation}
and
\begin{equation}\label{eq:eta-4}
\eta_-=\frac{r_0^4\left\{4Ma^2(r_0-h)e^{-\frac{h}{r_0}}-\left[r_0^2+M(-3r_0+h)e^{-\frac{h}{r_0}}\right]^2\right\}}{a^2\left[r_0^{2}-M(r_0+h)e^{-\frac{h}{r_0}}\right]^2},
\end{equation}
respectively.
Equations~(\ref{eq:xi-4}) and (\ref{eq:eta-4}) are the same as Eqs. (22) and (23) in Ref.~\cite{Amir:2016cen}.

\subsubsection{A rotating black hole considered by Tinchev~\cite{Tinchev:2015apf}}
Tinchev suggested a rotating regular black hole with a function
\begin{eqnarray}
m(r)=Me^{-\frac{j}{r^{2}}},
\end{eqnarray}
where $j$ is a constant, and Tinchev calculated the contour of the shadow of the black hole~\cite{Tinchev:2015apf}.
From Eqs.~(\ref{eq:xi-}) and~(\ref{eq:eta-}), we get 
\begin{equation}\label{eq:xi-5}
\xi_-=\frac{4Mr_0^4e^{-\frac{j}{r_0^2}}-\left[r_0^3+M(r_0^2+2j)e^{-\frac{j}{r_0^2}}\right](r_0^{2}+a^{2})}{a\left[r_0^{3}-M(r_0^2+2j)e^{-\frac{j}{r_0^2}}\right]}
\end{equation}
and
\begin{eqnarray}\label{eq:eta-5}
\eta_-
&=&\frac{r_0^4}{a^2\left[r_0^{3}-M(r_0^2+2j)e^{-\frac{j}{r_0^2}}\right]^2} \nonumber\\
&&\times \left\{4Ma^2r_0(r_0^2-2j)e^{-\frac{j}{r_0^2}} \right. \nonumber\\
&&\left. -\left[r_0^3-M(3r_0^2-2j)e^{-\frac{j}{r_0^2}}\right]^2\right\}.
\end{eqnarray}

Equation~(\ref{eq:xi-5}) is equal to $\xi_-$ in Eq. (13) in Ref.~\cite{Tinchev:2015apf}
while Eq.~(\ref{eq:eta-5}) is not equal to $\eta_-$ in Eq. (13) in Ref.~\cite{Tinchev:2015apf}, 
which is given by, in our notation,
\begin{equation}\label{eq:eta-5b}
\eta_-
=\frac{2\left\{ (r_0^{2}+a^{2}) \left[ (r_0m_0)'^{2}+r_0^{2} \right] -4r_0^{2}m_0(r_0m_0)' \right\}}{[(r_0m_0)'-r_0]^{2}}.
\end{equation}
When $j$ vanishes, the rotating regular black hole is the Kerr black hole.  
For $j=0$, Eq.~(\ref{eq:eta-5}) is equal to $\eta_-$ in the Kerr black hole spacetime obtained as Eq.~(\ref{eq:eta-1b})
while Eq. (\ref{eq:eta-5b}) becomes 
\begin{equation}\label{eq:eta-5c}
\eta_-
=\frac{2 \left[ (r_0^2+a^2)(r_0^2+M^2)-4M^2r_0^2 \right] }{(r_0-M)^2}
\end{equation}
and it is not the same as Eq.~(\ref{eq:eta-1b}).

\subsection{A rotating black hole considered by Atamurotov, Ghosh, and Ahmedov~\cite{Atamurotov:2015xfa}}
Atamurotov \textit{et al.} investigated the contour of the shadow of a rotating black hole~\cite{Atamurotov:2015xfa}
and they claimed that the rotating black hole was a rotating black hole obtained in Ref.~\cite{CiriloLombardo:2004qw}.
The rotating black hole in Ref.~\cite{Atamurotov:2015xfa} has the function $m(r)$ given by
\begin{eqnarray}
m(r)=M-\frac{K^2(r)}{2r}, 
\end{eqnarray}
where $K(r)$ is a function with respective to $r$. 
Please see Refs.~\cite{Atamurotov:2015xfa} and \cite{CiriloLombardo:2004qw} carefully.

From Eqs.~(\ref{eq:xi-}) and~(\ref{eq:eta-}), we obtain
\begin{equation}\label{eq:xi-6}
\xi_-=\frac{2r_0(2Mr_0-K_0^2)-(r_0+M-K_0K_0')(r_0^2+a^2)}{a(r_0-M+K_0K_0')}
\end{equation}
and
\begin{eqnarray}\label{eq:eta-6}
\eta_-
&=&\frac{r_0^2 }{a^{2}(r_0-M+K_0K_0')^{2}} \nonumber\\
&&\times \left\{ 4a^2 \left[ r_0(M+K_0K_0')-K_0^2\right] \right. \nonumber\\
&&\left.-\left[r_0(r_0-3M-K_0K_0')+2K_0^2 \right]^2 \right\},
\end{eqnarray}
where $K_0$ and $K_0'$ are defined as $K_0\equiv K(r_0)$ and $K_0'\equiv K'(r_0)$, respectively.
Equation~(\ref{eq:xi-6}) is equal to Eq.~(24) in Ref.~\cite{Atamurotov:2015xfa}
while Eq.~(\ref{eq:eta-6}) is not equal to $\eta_-$ obtained by Atamurotov \textit{et al.} as seen Eq. (25) in Ref.~\cite{Atamurotov:2015xfa}.
Atamurotov \textit{et al.} obtained $\eta_-$ as, in our notation, 
\begin{equation}\label{eq:eta-6b}
\eta_-
=\frac{16\Delta_0 r_0^{2}a^{2}-(r_0^{2}+a^{2})\Delta_0'+4r\Delta_0+\Delta_0'a}{(\Delta_0'a)^{2}},
\end{equation}
where $\Delta_0'$ is defined by $\Delta_0'\equiv \Delta'(r_0)$.
From a dimensional analysis, we notice that Eq. (25) in Ref.~\cite{Atamurotov:2015xfa} or Eq.~(\ref{eq:eta-6b}) should be modified.

\subsection{A rotating black hole considered by Modesto and Nicolini~\cite{Modesto:2010rv}}
Modesto and Nicolini applied the Newman-Janis algorithm to 
a noncommutative geometry inspired Reissner-Nordstr\"om solution obtained in Ref.~\cite{Ansoldi:2006vg}
and obtained a rotating black hole metric~\cite{Modesto:2010rv}.
The function $m(r)$ of the rotating black hole spacetime is given by 
\begin{eqnarray}
m(r)=n(r)-\frac{q^{2}(r)}{2r},
\end{eqnarray}
where $n(r)$ and $q(r)$ are functions with respect to $r$. 
See Eqs.~(37), (45) and (46) in Ref.~\cite{Modesto:2010rv} or Eq.~(1) 
in Ref.~\cite{Sharif:2016znp}.~\footnote{$n(r)$, in our notation, 
is $m(r)$ in a notation in Refs.~\cite{Sharif:2016znp,Modesto:2010rv}.}

The contour of the shadow of the rotating black hole was investigated by Sharif and Iftikhar~\cite{Sharif:2016znp}.
From Eqs.~(\ref{eq:xi-}) and~(\ref{eq:eta-}), we obtain $\xi_-$ and $\eta_-$ as  
\begin{eqnarray}\label{eq:xi-7a}
\xi_-
&=&\frac{1}{a(r_0-n_0-n_0'r_0+q_0q_0')} \left[ 2r_0(2n_0r_0-q_0^2) \right.  \nonumber\\
&&\left. -(r_0+n_0+n_0'r_0-q_0q_0')(r_0^2+a^2) \right]
\end{eqnarray}
and 
\begin{eqnarray}\label{eq:eta-7a}
\eta_-
&=&\frac{r_0^2}{a^2(r_0-n_0-n_0'r_0+q_0q_0')^2} \nonumber\\
&&\times \left\{ 4a^2\left[ r_0(n_0-n_0'r_0+q_0q_0')-q_0^2 \right] \right.  \nonumber\\
&&\left. -\left[ r_0(r_0-3n_0+n_0'r_0-q_0q_0')+2q_0^2 \right]^2  \right\}, \nonumber\\
\end{eqnarray}
respectively, where $n_0$, $n'_0$, $q_0$, and $q_0'$ are defined as 
$n_0\equiv n(r_0)$, $n'_0 \equiv n'(r_0)$, $q_0 \equiv q(r_0)$, and $q_0' \equiv q'(r_0)$, respectively.
We notice that Eqs.~(\ref{eq:xi-7a}) and (\ref{eq:eta-7a}) are not equal to Eqs.~(15) and (16) in Ref.~\cite{Sharif:2016znp}. 
Equations (15) and (16) in Ref.~\cite{Sharif:2016znp} are given by, in our notation, 
\begin{eqnarray}\label{eq:xi-7b}
\xi_-
&=&\frac{1}{a[n_0+r_0(n_0'-1)-q_0q_0']} \left[ n_0(a^{2}-3r_0^{2}) \right.  \nonumber\\
&&\left. +r_0(r_0^{2}+a^{2})(n_0'+1)+2q_0^{2}-q_0q_0'(r_0^{2}+a^{2}) \right] \nonumber\\
\end{eqnarray}
and
\begin{eqnarray}\label{eq:eta-7b}
\eta_-
&=&\frac{r_0^{2}}{a^{2}[n_0+r_0(n_0'-1)]^{2}} \left[ n_0r_0(4a^{2}-9n_0r_0+6r_0^{2}) \right.  \nonumber\\
&&-2n_0'r_0^{2}(2a^{2}+r_0^{2}-3n_0r_0) -r_0^{4}(n_0'^{2}+1) \nonumber\\
&&-4q_0^{2}(a^{2}+q_0^{2}-3n_0r_0+n_0'r_0^{2}+r_0^{2}) \nonumber\\
&&\left. -q_0'(4a^{2}+4q_0^{3}-6n_0q_0'r_0-2n_0'q_0r_0^{2}-q_0r_0+2q_0r_0^{2}) \right], \nonumber\\
\end{eqnarray}
respectively.
From a dimensional analysis, 
we notice that Eqs.~(15) and (16) in Ref.~\cite{Sharif:2016znp} or Eqs.~(\ref{eq:xi-7b}) and (\ref{eq:eta-7b}) should be modified.

\section{Summary}
We have obtained a formula for the contour of the shadow of rotating black holes generated by the Newman-Janis algorithm. 
We have applied the formula to two new examples of the contour of the shadow of rotating black holes.   
We notice the shadows is very similar to the shadow of the Kerr-Newman black hole. 
By using the formula, we have examined $\xi_-$ and $\eta_-$ of the Kerr-Newman black hole and rotating regular black holes and the other rotating black holes.
The formula would help to categorize the many known results of the shadow of rotating black holes and to obtain new examples of black hole shadows.

\section*{Acknowledgements}
The author thanks R.~A.~Konoplya for his useful comment. 
This research was supported in part by the National Natural Science Foundation of China under Grant No. 11475065,
the Major Program of the National Natural Science Foundation of China under Grant No. 11690021.
\appendix

\section{Celestial Coordinates $\alpha$ and $\beta$}
In this Appendix, we show the calculation for the celestial coordinates $\alpha$ and $\beta$~\cite{Bardeen:1973tla,Chandrasekhar:1983}.
We can express the line element~(\ref{eq:line_1}) 
in the asymptotically-flat, stationary, and axisymmetric black hole spacetime in the Boyer-Lindquist coordinates as follows:
\begin{eqnarray}
ds^{2}
= -e^{2\nu}dt^{2} +e^{2\psi} \left( d\phi-\omega dt \right)^{2} +e^{2\chi}dr^{2} +\rho^{2}d\theta^{2}, \qquad 
\end{eqnarray}
where
\begin{eqnarray}
&&e^{\nu} \equiv \sqrt{\frac{\rho^{2}\Delta}{\Sigma}}, \\
&&e^{\psi} \equiv \sqrt{\frac{\Sigma \sin^{2}\theta}{\rho^{2}}}, \\
&&\omega \equiv \frac{2m(r)ar}{\Sigma}, \\
&&e^{\chi} \equiv \sqrt{\frac{\rho^{2}}{\Delta}}.
\end{eqnarray}
The inverse of the metric tensors are obtained as
\begin{eqnarray}
&&g^{tt} = -e^{-2\nu}, \\
&&g^{t \phi} =-\omega e^{-2\nu}, \\
&&g^{\phi \phi} =-\omega^{2}e^{-2\nu} +e^{-2\psi}, \\
&&g^{rr} = e^{-2\chi}, \\
&&g^{\theta \theta} = \rho^{-2}.
\end{eqnarray}
We use a tetrad frame by the basis-vectors~$e_{(\beta)\alpha}dx^{\alpha}$ 
and the contravariant basis-vectors~$e_{(\beta)}^{\alpha}\partial_{\alpha}$:
\begin{eqnarray}
&&e_{(0)\alpha}dx^{\alpha}=-e^{\nu}dt, \\
&&e_{(1)\alpha}dx^{\alpha}=-\omega e^{\psi}dt +e^{\psi}d\phi, \\
&&e_{(2)\alpha}dx^{\alpha}=e^{\chi}dr, \\
&&e_{(3)\alpha}dx^{\alpha}=\rho d\theta,
\end{eqnarray}
and 
\begin{eqnarray}
&&e_{(0)}^{\alpha} \partial_{\alpha} = e^{-\nu} \partial_{t} +\omega e^{-\nu} \partial_{\phi}, \\
&&e_{(1)}^{\alpha} \partial_{\alpha} = e^{-\psi} \partial_{\phi}, \\
&&e_{(2)}^{\alpha} \partial_{\alpha} = e^{-\chi} \partial_{r}, \\
&&e_{(3)}^{\alpha} \partial_{\alpha} = \rho^{-1} \partial_{\theta}.
\end{eqnarray}
The tetrad components of the 4-momentum of the photon  $p^{(\beta)}$ are obtained as
\begin{eqnarray}
p^{(0)}
&=&-p_{(0)}=-e^{\nu}p^{t} \nonumber\\
&=&-e^{\nu}\left[ (E-\omega L) e^{-2\nu} \right] = (\omega L -E) e^{-\nu}, \\
p^{(1)}
&=&p_{(1)}=e^{\psi} \left( -p^{t}\omega+ p^{\phi} \right) =Le^{-\psi}, \\
p^{(2)}
&=&p_{(2)}=e^{\chi}p^{r},\\
p^{(3)}
&=&p_{(3)}=\rho p^{\theta},
\end{eqnarray}

In an asymptotic region $r \rightarrow \infty$, we obtain
\begin{eqnarray}
&&e^{\nu} \rightarrow 1, \\
&&e^{\psi} \rightarrow r \sin\theta, \\
&&\omega \rightarrow \frac{2Ma}{r^{3}}, \\
&&e^{\chi} \rightarrow 1.
\end{eqnarray}
Thus, as $r \rightarrow \infty$, the tetrad components of the 4-momentum  of the photon $p^{(\beta)}$ are given by
\begin{eqnarray}
&&p^{(0)} \rightarrow -E, \\
&&p^{(1)} \rightarrow \frac{L}{r \sin \theta}, \\
&&p^{(2)} \rightarrow p^{r},\\
&&p^{(3)} \rightarrow r p^{\theta}.
\end{eqnarray}

The contour of the shadow of an asymptotically-flat, stationary, and axisymmetric black hole seen by an observer 
at a large distance from the black hole with an inclination angle $\theta_i$ is expressed 
by celestial coordinates $\alpha$ and $\beta$~\cite{Bardeen:1973tla}.
The celestial coordinates $\alpha$ and $\beta$ are defined by
\begin{eqnarray}
\alpha 
\equiv \lim_{r\rightarrow \infty} \left( -\frac{rp^{(1)}}{p^{(0)}} \right) 
= \frac{\xi}{\sin \theta_i}
\end{eqnarray}
and
\begin{eqnarray}
\beta
\equiv \lim_{r\rightarrow \infty} \left( -\frac{rp^{(3)}}{p^{(0)}} \right) 
= \sigma_{\theta} \sqrt{\frac{\Theta}{E^{2}}},
\end{eqnarray}
respectively.

\section{Comment on Li and Bambi~\cite{Li:2013jra}}
In this short Appendix, we comment on Ref.~\cite{Li:2013jra}.
Li and Bambi claimed that $\xi_-$ and $\eta_-$ of the rotating Bardeen black hole and the rotating Hayward black hole are complicated 
and they did not show the explicit forms of $\xi_-$ and $\eta_-$ in Ref.~\cite{Li:2013jra}. 
$\xi_-$ and $\eta_-$, however, are not complicated so much
as we showed them in Eqs.~(\ref{eq:xi-8}) and (\ref{eq:eta-8}) for the rotating Bardeen black hole 
and in Eqs.~(\ref{eq:xi-3}) and (\ref{eq:eta-3}) for the rotating Hayward black hole.

We note that we should read Eq.~(3.6) in Ref.~\cite{Li:2013jra}, in our notation,
\begin{equation}
\left. R'(r)\right|_{r=r_{0}}
=4r_0^3+2(a^2-\xi^2-\eta)r_0+2m_0[\eta+(\xi-a)^2]
=0 
\end{equation}
as 
\begin{equation}
\left. R'(r)\right|_{r=r_{0}}
=4r_0^3+2(a^2-\xi^2-\eta)r_0+2m_0[\eta+(\xi-a)^2]f_0
=0 
\end{equation}
which was obtained as Eq.~(2.17) in Tsukamoto \textit{et al.}~\cite{Tsukamoto:2014tja}.

\end{document}